# New Approach for CCA2-Secure Post-Quantum Cryptosystem Using Knapsack Problem

Roohallah Rastaghi

r.rastaghi59@gmail.com

*Abstract*— Chosen-ciphertext security, which guarantees confidentiality of encrypted messages even in the presence of a decryption oracle, has become the *de facto* notion of security for public-key encryption under active attack. In this manuscript, for the first time, we propose a new approach for constructing post-quantum cryptosystems secure against adaptive chosen ciphertext attack (CCA2-secure) in the standard model using the knapsack problem. The *computational* version of the knapsack problem is *NP-hard*. Thus, this problem is expected to be difficult to solve using quantum computers. Our construction is a precoding-based encryption algorithm and uses the knapsack problem to perform a permutation and pad random fogged data to the message bits. Compared to other approaches in use today, our approach is more efficient and its CCA2 security in quantum environment can be reduced in the standard model to the assumption that the knapsack problem is intractable. Furthermore, we show that our approach is a general paradigm and can be applying to any (post-quantum) trapdoor one-way function candidate.

## I. Introduction

THE notion of public-key cryptography was introduced by Diffy and Hellman [18] and the ultimate goal of public-key encryption is the production of a simple and efficient encryption scheme that is provably secure in a strong security model under a weak and reasonable computational assumption. The accepted notion for the security of a public-key encryption scheme is semantically secure against adaptive chose ciphertext attack (i.e. IND-CCA2) [37]. In this scenario, the adversary has seen the *challenge ciphertext* before having access to the decryption oracle. The adversary is not allowed to ask the decryption of the challenge ciphertext, but can obtain the decryption of any relevant cryptogram (even modified ones based on the challenge ciphertext). A cryptosystem is said to be CCA2-secure if the cryptanalyst fails to obtain any partial information about the plaintext relevant to the challenge cryptogram.

In order to design CCA2-secure cryptosystems, a lot of successes were reached but these successes tend to fall into two categories: the production of very efficient encryption schemes with security proofs in random oracle models, and the production of less efficient encryption schemes with full proofs of security in standard models. The ultimate prize has yet to be claimed. The random oracle model was introduced by Bellare and Rogaway [5]. In this model, hash functions are considered to be ideal, i.e. perfectly random. We stress that although a proof in the random oracle model has a certain value it is still only a heuristic security argument for any implementation of the scheme. In particular, there exist cryptographic schemes that are provably secure in the random oracle model yet that are insecure with *any* possible standard model instantiation of the hash function [10]. The approaches for constructing encryption schemes secure in the standard model (i.e., without the use of heuristics such as random oracles) tend to fall several categories [17].

- The first approach is to use a "double-and-add" technique, in which a message is encrypted twice and a checksum value is added to the ciphertext. We can divide these schemes in two categories: The NIZK Schemes [33] and the Cramer-Shoup encryption Scheme [16].
- Signature and identity-based schemes are another approach for constructing IND-CCA2 scheme. The Canetti-Halevi-Katz transform [11] and the Dolev-Dwork-Naor scheme [19] are the most important cases in this area.
- Extracting plaintext awareness is the third technique. Two methods are used: plaintext awareness via key registration and using Extractors [4, 21].

The cryptosystems were introduced with the above techniques are based on assumptions related to the integer factoring (IF) and discrete logarithm (DL) problems, such as: decisional quadratic residuosity (DQR) assumption, Paillier's decisional composite residuosity (DCR) assumption, RSA assumption, decisional Diffie-Hellman (DDH) assumption, Computational Diffie-Hellman (CDH) assumption, Bilinear Computational Diffie-Hellman (BCDH) assumption, etc. In 1994, Shor [42] conceived a polynomial-time algorithm which solves IFP and DLP using quantum computers. Therefore, these schemes cannot be used to provide security protections any longer and public-key cryptosystems secure in quantum computing environments need to be developed. Public-key cryptosystems that can resist these emerging attacks are called quantum resistant or post-quantum cryptosystems.

*A. Related work*

There are mainly four classes of public-key cryptography that are believed to resist quantum attacks [6].
- Lattice-based cryptography
- Code-based cryptography
- Hash-based digital signature schemes
- Multivariate cryptography

Each approach has unique advantages and disadvantages. It has been shown that many of these post-quantum cryptosystems are not CCA2-secure [7, 23] and most of the cryptosystems have been proposed to deal with CCA2 adversaries are in the random oracle model. In recent years, Lattice-based and Code-based cryptography have been further considered in the design of post-quantum CCA2-secure cryptosystem. Here we present a summary of the work done in each case.

- *Lattice-based cryptography.* For the first time, Peikert and Waters [35] propose a CCA1-secure cryptosystem based on the worst-case hardness of lattice problems using the notion of lossy All-But-One (ABO) trapdoor functions. Subsequently, Peikert [36] showed how to construct a correlation-secure trapdoor function family from the learning with errors (LWE) problem, and used it within the Rosen-Segev scheme [40] to obtain another lattice-based CCA1-secure scheme. Unfortunately, the latter scheme suffers from long public-key and ciphertext length of $\Omega(n^2)$ in the security parameter $n$, even if applied in the Ring-LWE setting [43]. Recently, Micciancio and Peikert [31] give new methods for generating simpler, tighter, faster and smaller trapdoors in cryptographic lattices to achieve a CCA1-secure cryptosystem. Their methods involve a new kind of trapdoor, and include specialized algorithms for inverting LWE, randomly sampling SIS preimages, and securely delegating trapdoors. More constructions of IND-CCA2 secure lattice-based encryption schemes can be obtained by using the lattice-based selective-ID secure identity-based encryption (IBE) schemes of [1, 2, 3, 39] within the generic construction of [8], and a one-time signature or commitment scheme. Recently, Steinfeld et al. [43] introduced the first CCA2 secure variant of the NTRU in the standard model with a provable security from worst-case problems in *ideal lattices*. They construct a CCA1-secure scheme using the lossy trapdoor function framework of Peikert and Waters, which they generalize to the case of $(k-1)$-of-$k$-correlated input distributions.

    For full CCA2 security and non-malleability, all the above schemes suggest using a strongly unforgeable one-time signature scheme, a message authentication code (MAC) or commitment scheme for transfer CCA1-secure schemes to a CCA2-secure one. So, all of these approaches require *separate encryption* and *checksum operations*. Hence, the resulting encryption schemes require two expensive calculations.

- *Code-based cryptography.* The first cryptosystem based on the hardness of decoding random linear codes was presented in 1978 by McEliece [27]. The first CCA2-secure variant of McEliece scheme in the random oracle model is the conversions of Kobara and Imai [26]. Recently, Cayrel et al. [13] present efficient implementations of McEliece variants using quasi-dyadic codes. They provide secure parameters for a classical McEliece encryption scheme based on quasi-dyadic generalized Srivastava codes, and convert their scheme to a CCA2-secure one in the random oracle model by applying the Fujisaki-Okamoto transform. For the first time, Döttling et al. [20] showed how we can construct CCA2-secure McEliece variant cryptosystem using correlated products in the standard model. They construct IND-CPA verifiable $k$-repetition McEliece scheme by modifying Nojima et al. [34] randomized version of the McEliece with Bernoulli error distribution. Their scheme obtains CCA2-security in the standard model by using one-time strongly unforgeable signature scheme.

All the above concrete constructions of lossy trapdoor function and correlated inputs function are based on *decisional* assumptions [25, 28]. It is widely believed that computational assumptions are more standard than their decisional versions. Furthermore, IND-CCA2 for almost all proposed post-quantum cryptosystems is based on strongly unforgeable one-time signature schemes (or MACs). The drawback of the one-time signature schemes is that one key-pair consisting of a secret signature key and a public verification key can only be used to sign and verify a single document. This is inadequate for most applications [6]. Moreover, both the double-and-add schemes and the identity-based schemes require separate encryption and checksum operations. Hence, the resulting encryption schemes require two *expensive* calculations.

*B. Motivation*

Challenges in the post-quantum cryptography include efficiency, reliability and usability. Motivated by the task of constructing a simple and efficient public-key encryption scheme which is *post-quantum CCA2-secure* in the *standard model*, we initiate the study of *NP-hard* problems for constructing such schemes. It is believed that the quantum computer cannot solve NP-hard problems. The knapsack problem has proven to be NP-complete. The *computational* version of the knapsack problem is *NP-hard*, thus this problem is expected to be one of the difficult problems against the quantum computer. The knapsack scheme is also attractive at the standpoint of encryption and decryption speed. Only integer additions are required for the encryption in the knapsack scheme. This feature confers a practical advantage that the encryption can be lightly implemented on small IC chips.

*C. Overview*

We introduce a novel approach for constructing CCA2-secure post-quantum cryptosystem in the standard model based on the knapsack problem. Our construction is a randomized precoding-based algorithm and uses the knapsack problem to perform a permutation and pad random fogged data to the message bits. For this reason, we firstly choose a random binary string $\{x_1, x_2, \ldots, x_n\} \in_r \{0,1\}^n$ and then use it to divide message bits into several blocks with equal length $v$. Then, some random fogged blocks (RFBs) with equal length $s$ are padded to the message blocks. Encoding information includes *position* of the message (and RFB) blocks and the *lengths* of them are encrypted with a secure knapsack-based PKC and send to the receiver. Since encryption and decryption in the knapsack-based cryptosystems are very fast, compared to other approaches, our method is very efficient. The main novelty is that we introduce a heuristic encoding algorithm and correlate the encoding parameters to the knapsack problem. Thus, the CCA2 adversary for extracting the message from *challenge cryptogram* must first recover encoding parameters from the knapsack-based cryptosystem. Our proof makes direct use of the fact that the underlying primitive is a trapdoor one-way permutation, rather than the knapsack problem and the scheme's consistency check can be directly implemented by the simulator without having access to some external gap-oracle (as in [8, 9, 24]) or using other extrinsic rejection techniques (such as "hash proof systems" [1, 5, 16] or "twinning" [12]). Thus, our proof technique is fundamentally different from all known approaches to obtain CCA security.

The rest of this manuscript is organized as follows: In the following section, we briefly explain some mathematical background and definitions. Then, in Section 3, we introduce our proposed scheme. Performance and security analysis of this cryptosystem will be discussed in Section 4 and finally, in section 5, we will give a general form of the proposed scheme that can be applying to any (post-quantum) trapdoor one-way function candidate.

## II. PRELIMINARY

### A. Notation

We will use standard notation. If $X$ is a string, then $|X|$ denotes its length. If $k \in \mathbb{N}$ then $\{0,1\}^k$ denote the set of $k$-bit strings, $1^k$ denote a string of $k$ ones and $\{0,1\}^*$ denote the set of bit strings of finite length. $y \leftarrow x$ denotes the assignment to $y$ of the value $x$. For a set $S$, $s \leftarrow S$ denote the assignment to $s$ of a uniformly random element of $S$. For a deterministic algorithm $A$, we write $x \leftarrow A^O(y, z)$ to mean that $x$ is assigned the output of running $A$ on inputs $y$ and $z$, with access to oracle $O$. If $A$ is a probabilistic algorithm, we may write $x \leftarrow A^O(y, z, R)$ to mean the output of $A$ when run on inputs $y$ and $z$ with oracle access to $O$ and using the random coins $R$. If we do not specify $R$ then we implicitly assume that the coins are selected uniformly at random from $\{0,1\}^\infty$. This is denoted $x \leftarrow A^O(y, z)$. We denote by $\Pr[E]$ the probability that the event $E$ occurs. If $a$ and $b$ are two strings of bits, we denote by $a || b$ their concatenation. The bit-length of $a$ denoted by $\text{Len}(a)$, $\text{Lsb}_{x_1}(a)$ means the right $x_1$ bits of $a$ and $\text{Msb}_{x_2}(a)$ means the left $x_2$ bits of $a$.

### B. Public-Key Encryption Schemes

A public-key encryption scheme is defined as follows:

*Definition 1:* (Public-key encryption). A public-key encryption scheme (PKE) is a triple of probabilistic polynomial time (PPT) algorithms $(\text{Gen}, \text{Enc}, \text{Dec})$ such that:

- $\text{Gen}$ is a probabilistic polynomial-time key generation algorithm which takes a security parameter $1^k$ as input and outputs a public-key $pk$ and a secret-key $sk$. We write $(pk, sk) \leftarrow G(1^k)$. The public-key specifies the message space $\textbf{MsgSp}(pk)$ and the ciphertext space $\textbf{CiphSp}(sk)$.
- $\text{Enc}$ is a (possibly) probabilistic polynomial-time encryption algorithm which takes as input a public key $pk$, a $m \in \textbf{MsgSp}(pk)$ and random coins $r$, and outputs a ciphertext $C \in \textbf{CiphSp}(pk)$. We write $\text{Enc}(pk, m; r)$ to indicate explicitly that the random coins $r$ is used and $\text{Enc}(pk, m)$ if fresh random coins are used.
- $\text{Dec}$ is a deterministic polynomial-time decryption algorithm which takes as input a secret-key $sk$ and a ciphertext $C \in \textbf{CiphSp}(pk)$, and outputs either a message $m \in \textbf{MsgSp}(pk)$ or an error symbol $\bot$. We write $m \leftarrow \text{Dec}(C, sk)$.
- (Completeness) For any pair of public and secret-keys generated by $\text{Gen}$ and any message $m \in \textbf{MsgSp}(pk)$ it holds that $\text{Dec}(sk; \text{Enc}(pk, m; r)) = m$ with overwhelming probability over the randomness used by $\text{Gen}$ and the random coins $r$ used by $\text{Enc}$.

*Semantic security* (a.k.a. indistinguishability) against adaptive chosen ciphertext attacks (IND-CCA2) is the strongest known notion of security for the public-key encryption schemes. This notion of security simply means that a cryptogram should not reveal any useful information about the message. This type of attack is the most powerful attack, which is defined by Rackoff and Simon [37]. In this scenario, the adversary has seen the *challenge cryptogram* before having access to the decryption oracle. The adversary is not allowed to ask the decryption of the challenge cryptogram, but can obtain the decryption of any relevant cryptogram (even modified ones based on the challenge cryptogram). A cryptosystem is said to be IND-CCA2 secure if the cryptanalyst fails to obtain any partial information about the plaintext relevant to the challenge cryptogram. This notion is known to suffice for many applications of encryption in the presence of active attackers including secure P2P transmission, secure communication, auctions, voting schemes, and many others. Indeed, CCA security is now widely accepted as the standard security notion for public-key encryption schemes. In recent years, attempt to design cryptosystems secure against adaptive chosen ciphertext attack is increased, but no efficient cryptosystems in the standard model were introduced.

*Definition2* (CCA2-security). A public-key encryption scheme PKE is secure against adaptive chosen-ciphertext attacks (i.e. IND-CCA2) if the advantage of any two-stage PPT adversary $A = (A_1, A_2)$ in the following experiment is negligible in the security parameter $k$:

$$\text{Exp}_{\text{PKE},A}^{cca2}(k):$$
$$(pk, sk) \leftarrow \text{Gen}(1^k)$$
$$(m_0, m_1, state) \leftarrow A_1^{\text{Dec}(sk,.)}(pk) \text{ s.t. } |m_0| = |m_1|$$
$$b \leftarrow \{0,1\}$$
$$C^* \leftarrow \text{Enc}(pk, m_b)$$
$$b' \leftarrow A_2^{\text{Dec}(sk,.)}(C^*, state)$$
$$\text{If } b = b' \text{ return } 1, \text{ else return } 0.$$

The attacker may query a decryption oracle with a ciphertext $C$ at any point during its execution, with the exception that $A_2$ is not allowed to query $\text{Dec}(sk,.)$ with $C^*$. The decryption oracle returns $b' \leftarrow A_2^{\text{Dec}(sk,.)}(C^*, state)$. The attacker wins the game if $b = b'$ and the probability of this event is defined as $\Pr[\text{Exp}_{\text{PKE},A}^{cca2}(k)]$. We define the advantage of $A$ in the experiment as

$$\text{Adv}_{\text{PKE},A}^{\text{IND-CCA-2}}(k) = \left| \Pr[\text{Exp}_{\text{PKE},A}^{cca2}(k) = 1] - \frac{1}{2} \right|.$$

### A. Knapsack-based public-key cryptosystem

Knapsack public-key encryption schemes are based on the subset sum problem. The subset sum problem is to find the solution $(x_1, x_2, \ldots, x_n) \in \{0,1\}^n$ such that $C = a_1 x_1 + a_2 x_2 + \ldots + a_n x_n$ for given positive integers $a_1, a_2, \ldots, a_n$ and $C$ which is the sum of a subset of the $a_i$. This problem is known to be *NP-hard* [30, 32]. Thus, the subset sum problem is expected to be hard to solve using quantum computers. The knapsack scheme is also attractive at the standpoint of light-weight encryption. In the public-key schemes using DLP or IFP, the large modulo exponentiations are required for the encryption and decryption. This operation is a little bit heavy to implement on small IC chips. On the other hand, only integer additions are required for the encryption in the knapsack scheme. This feature confers a practical advantage that the encryption can be lightly implemented on small IC chips and can be performed in much shorter time than the schemes using DLP or IFP. These reasons described above have motivated us to investigate knapsack-based public-key cryptosystems.

The first knapsack-based PKC was introduced by Merkle and Hellman [29]. Since its proposal, knapsack-based PKC have been widely studied and many knapsack-based PKCs were developed. Nevertheless, many knapsack cryptosystems were shown insecure against some known attacks, such as low density attack [14], Shamir's attack [41], and lattice basis reduction algorithms [38], etc. The basic idea is to select an instance of the subset sum problem that is easy to solve, and then disguise it as an instance of the general subset sum problem, which is hopefully difficult to solve. The original knapsack set can serve as the private key, while the transformed knapsack set serves as the public key [30].

## I. THE PROPOSED CRYPTOSYSTEM

In this section, we introduced our proposed encryption scheme. Our scheme is a precoding-based algorithm that uses knapsack problem for performing a permutation and pad random fogged data to the message.

### A. The Proposed Idea

Let we can decide to encrypt message $m \in \{0,1\}^n$ with $\text{Len}(m) = n$. For perform a randomized encoding to the message, we firstly choose a random binary vector $X = (x_1, x_2, \ldots, x_l)$ with hamming weigh $h$ where $h = \sum_{i=1}^{l} x_i$. We will call this vector *cargo (carrier) vector* which carries the encoding information. If $h \nmid n$, then $n/h$ is not an integer and we pad a random binary string (RBS) with length $h \cdot \lceil n/h \rceil - n$ to the right of $m$. Otherwise, $n/h$ is an integer and we don't need to pad a RBS to the right of $m$. In each case, we can divide $m$ into $h$ blocks $d_1 \| d_2 \| \ldots \| d_h$ with binary length $v = \lceil n/h \rceil$ where

$d_h = \text{Lsb}_m(n-(h-1).\lceil n/h \rceil) \| \text{RBS}$. Thus, if $h \mid n$, then $\text{RBS} = \phi$ (the empty set) and $d_h = \text{Lsb}_m(n-(h-1).\lceil n/h \rceil)$, else, RBS is a random block with binary length $h.\lceil n/h \rceil - n$ and we have $d_h = \text{Lsb}_m(n-(h-1).\lceil n/h \rceil) \| \text{RBS}$.

We perform a random permutation and pad some random fogged blocks (RFBs) with equal binary length $s$ into the message blocks $d_i, 1 \le i \le h$ using permutation function $f: \{0,1\} \times \{0,1\}^v \to \{0,1\}^v \times \{0,1\}^s$, that can be defined as follows:

$$f(x_i, d_i) = d_i' = \begin{cases} d_i & \text{if } x_i = 1 \\ \text{RFB} & \text{if } x_i = 0 \end{cases}.$$

Notice that in order to prevent excessive increase in the message length, we can choose $s$ small enough. The message $m' = (d_1' \| d_2' \| ... \| d_l')$ is called encoded message. We summarize encoding process in algorithm1.

---

**Algorithm1**: Permutation and random data padding

Input: $m \leftarrow (m_1, m_2, ..., m_n)$ and random binary vector $X = (x_1, x_2, ..., x_l)$

Output: Encoded message $m' = (d_1' \| d_2' \| ... \| d_l')$.

*SETUP:*

1. If $h \mid n$ then $v \leftarrow n/h$;

   else

   $v \leftarrow \lceil n/h \rceil$ and choose a RBS with $\text{Len}(h.\lceil n/h \rceil - n)$

   and $m \leftarrow (m_1, m_2, ..., m_n \underbrace{\square \text{ RBS }}_{h.\lceil n/h \rceil - n}))$;

2. Divide $m$ into $h$ blocks $(d_1 \| d_2 \| ... \| d_h)$ with $\text{Len}(d_i) = v$, $1 \le i \le h$

*PERMUTATION AND PADDING:*

1. Randomly chose integer $s$.
2. For $i = 1$ to $l$ do:

   If $x_i = 1$ then $d_i' \leftarrow d_{\sum_{j=1}^{i} x_j}$;

   else

   $d_i' \leftarrow$ RFB with binary length $s$.

   Return $m' = (d_1' \| d_2' \| ... \| d_l')$

---

We illustrate algorithm1 with small example. Suppose $m = (m_1, m_2, ..., m_{127})$ and $X = (0,1,0,1,1,0,1,0,1,1,1,0,1,1,1,0,1,0)$.

*SETUP:*

We have $\text{Len}(m) = n = 127$, $l = 18$, $h = \sum_{i=1}^{l} x_i = 11$. Since $11 \nmid 127$ so $v = \lceil n/h \rceil = 12$ and we must pad a random block with length $\text{Len}(h.\lceil n/h \rceil - n) = 5$ to the right of $m$. If we randomly chose $1,0,1,1,0$, we have $m = (m_1, m_2, ..., m_{127}, \underbrace{1,0,1,1,0}_{\text{Len}\{5\}})$. Since $h = 11$, the algorithm divides $m$ into 11 blocks with equal length $v = 12$. We have $m = (\underbrace{m_1, ..., m_{12}}_{d_1} \square \underbrace{m_{13}, ..., m_{24}}_{d_2} \square ... \square \underbrace{m_{121}, ..., m_{127}, 1, 0, 1, 1, 0}_{d_{11}})$.

Which $\text{Lsb}_m(n-(h-1).\lceil n/h \rceil) = \text{Lsb}_m(7) = m_{121}, ..., m_{127}$.

*PERMUTATION AND PADDING:*

Firstly, we choose random integer $s$, say $s = 4$. We have

$x_1 = 0$, thus $d_1' \leftarrow$ RFB#1 = $(0,1,1,0)$ where $(0,1,1,0)$ is randomly chosen by algorithm1.

$x_2 = 1$, thus $d_2' \leftarrow d_{\sum_{j=1}^{2} x_j} = d_1$.

$x_3 = 0$, thus $d'_3 \leftarrow \text{RFB\#2} = (1,0,1,0)$ where $(1,0,1,0)$ is randomly chosen by algorithm1.

.
.
.

$x_{17} = 1$, thus $d'_{17} \leftarrow d_{\sum_{j=1}^{17} x_j} = d_{11}$.

$x_{18} = 0$, thus $d'_{18} \leftarrow \text{RFB\#7} = (0,0,1,0)$ where $(0,0,1,0)$ is randomly chosen by algorithm1.

$l-h$ RFB blocks with equal length $s = 4$ are combined with the message blocks $d_i$, $1 \leq i \leq h$, to produce the encoded message $m' = (d'_1 \square d'_2 \square ... \square d'_l)$. In the final, algorithm output $m'$ as $m' = (\underbrace{0,1,1,0}_{d'_1} \square \underbrace{m_1,...,m_{12}}_{d'_2} \square \underbrace{1,0,1,0}_{d'_3} \square ... \square \underbrace{m_{121},...,m_{127},1,0,1,0}_{d'_{17}}$

$\square \underbrace{0,0,1,0}_{d'_{18}})$.

As we see, the *length* and *position* of the message blocks are correlated to the *number* and *position* of $x_i = 1, 1 \leq i \leq l$ respectively, and completely random.

### B. The proposed scheme

#### 1) Key Generation Algorithm

Suppose $\Pi = (\text{Gen}_{\text{Knap}}, \text{Enc}_{\text{Knap}}, \text{Dec}_{\text{Knap}})$ be a secure knapsack-based PKE and $(b_1, b_2, ..., b_l)$ be the secret key and $\{(a_1, a_2, ..., a_l), M\}$ be the public key, where $M > \sum_{i=1}^{l} a_i$ is a module.

#### 2) Encryption algorithm

To encrypt message $m \in \{0,1\}^n$, the sender do the following steps:

- Choose random cargo vector $X = (x_1, x_2, ..., x_l)$ with hamming weight $h$. Execute algorithm1 for generate encoded message $m' = (d'_1 \square d'_2 \square ... \square d'_l)$ from message $m$.

- Suppose $y$ be the decimal representation of $m'$. Compute:

$$C_1 = yr + h \quad , \quad C_2 = \text{Enc}_{\text{Knap}}(X) = \sum_{i=1}^{l} a_i . x_i$$

and send ciphertext $C = (C_1, C_2)$ to the receiver. Fig.1 shows the encryption process.

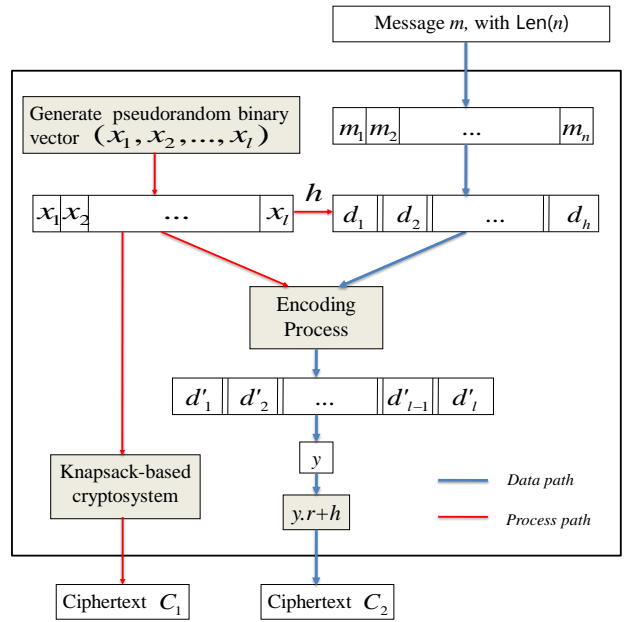

Fig.1: Encryption Process

*3) Decryption algorithm*

Receiver after receiving $C = (C_1, C_2)$, performs the following steps for extract message $m$:

- Compute random binary vector $X = (x_1, x_2, \ldots, x_l)$ as $X = \text{Dec}_{\text{Knap}}(C_2)$, $h = \sum_{i=1}^{l} x_i$ and $v = \lceil n/h \rceil$.
- Compute $y = (C_1 - h)/r$.
- Suppose $m'$ be the binary coded decimal (BCD) representation of $y$. Computes $s = \lfloor (\text{Len}(m') - n)/(l - h) \rfloor$ and reject the ciphertext if $s$ is not an integers (*consistency* check).
- The lengths and position of the message (and RFB) blocks are explicit, therefore, the receiver simply can separate RFB blocks from encoded message $m'$ and extract message blocks $d_i$, $1 \leq i \leq h$ with the following algorithm.

---

**Algorithm 2**: Extractor

Input: $X = (x_1, x_2, \ldots, x_l)$, $h = \sum_{i=1}^{l} x_i$, $s$ and $m'$

Output: Message $m = (d_1 \| d_2 \| \ldots \| d_h)$.

1. For $i = 1$ to $l$ do:

   If $x_i = 0$ then $m' \leftarrow \text{Lsb}_{\text{Len}(m') - s}(m')$;

   else

   $d_{\sum_{j=1}^{i} x_j} \leftarrow \text{Msb}_v(m')$ and $m' \leftarrow \text{Lsb}_{\text{Len}(m') - v}(m')$;

2. $m \leftarrow (d_1 \| d_2 \| \ldots \| d_{\sum_{j=1}^{l} x_j})$

3. If $h \nmid n$, then $m \leftarrow \text{Msb}_n(m)$ (remove right $(h \cdot \lceil n/h \rceil - n)$ bits of $m$).

Return "$m$"